\documentclass[12pt,3p,twocolumn]{elsarticle}
\usepackage{graphics}
\usepackage{amssymb}
\usepackage{natbib}
\biboptions{square}
\begin{document}

\begin{frontmatter}

\title{Low field magnetic response of the non-centrosymmetric superconductor YPtBi}

\author[wzi]{T. V. Bay}
\author[neel]{M. Jackson}
\author[neel]{C. Paulsen}
\author[psi]{C. Baines}
\author[psi]{A. Amato}
\author[stony]{T. Orvis}
\author[brookh,stony]{M. C. Aronson}
\author[wzi]{Y. K. Huang}
\author[wzi]{A. de Visser\corref{cor1}}
\ead{a.devisser@uva.nl}
\cortext[cor1]{Corresponding author}

\address[wzi]{Van der Waals - Zeeman Institute, University of Amsterdam, Science Park 904, 1098 XH Amsterdam, The Netherlands}
\address[neel]{Institut N\'{e}el and Universit\'{e} Joseph Fourier, CNRS, BP 166, 38042 Grenoble Cedex 9, France}
\address[psi]{Laboratory for Muon Spin Spectroscopy, PSI, CH-5232 Villigen, Switzerland}
\address[stony]{Department of Physics and Astronomy, Stony Brook University, Stony Brook NY 11794-3800 USA}
\address[brookh]{Condensed Matter Physics and Materials Science Department, Brookhaven National Laboratory, Upton NY 11973 USA}

\date{\today}

\begin{abstract}
The low-field magnetic response of the non-centrosymmetric superconductor YPtBi ($T_c =0.77$~K) is investigated. Ac-susceptibility and dc-magnetization measurements provide solid evidence for bulk superconductivity with a volume fraction of $\sim 70$~\%. The lower critical field is surprisingly small: $B_{c1} = 0.008$~mT $(T \rightarrow 0)$. Muon spin rotation experiments in a transverse magnetic field of 0.01~T show a weak increase of the Gaussian damping rate $\sigma_{TF}$ below $T_c$, which yields a London penetration depth $\lambda = 1.6 \pm 0.2~ \mu$m. The zero-field Kubo-Toyabe relaxation rate $\sigma_{KT}$ equals $0.129 \pm 0.004~\mu$s$^{-1}$ and does not show a significant change below $T_c$. This puts an upper bound of 0.04~mT on the spontaneous magnetic field associated with a possible odd-parity component in the superconducting order parameter.
\end{abstract}


%PACS 74.70.Dd, 76.75.+i, 74.25.Ha

\end{frontmatter}

\clearpage

\section{Introduction}

Recently, superconductivity was discovered in the equiatomic transition metal bismuthide YPtBi with a transition temperature  $T_c = 0.77$~K~\cite{Butch2011}. The superconducting state of YPtBi deserves a close examination because of two unusual aspects. Firstly, YPtBi crystallizes in the Half Heusler MgAgAs structure~\cite{Canfield1991} which lacks inversion symmetry, and, consequently, it is a non-centrosymmetric superconductor. The absence of an inversion center gives rise to an electric field gradient which creates an antisymmetric Rashba-type spin-orbit coupling, which in turn causes a splitting of the energy bands and Fermi surface. This has important consequences for the parity of the superconducting condensate, as it may give rise to the admixture of even and odd parity Cooper pair states, rather than to conventional spin-singlet states~\cite{Frigeri2004}. Non-centrosymmetric superconductors attract ample attention as test-case systems for research into unconventional superconducting phases~\cite{Bauer&Sigrist2012}. The second reason of interest is the possibility that YPtBi is a topological superconductor. Electronic structure calculations for a non-magnetic ternary Half Heusler compounds  predict a topologically non-trivial band structure, notably a substantial band inversion, due to strong spin-orbit coupling~\cite{Chadov2010,Lin2010,Feng2010}. Among the equiatomic platinumbismuthides, especially YPtBi, LaPtBi and LuPtBi have a strong band inversion, which makes them promising candidates for 3D topological insulating or topological semimetallic behaviour. A topological insulator has the intriguing property that its interior is an insulator, while the surface harbors metallic states that are protected by topology~\cite{Hasan&Kane2010,Qi&Zhang2010}. Indeed, YPtBi~\cite{Butch2011,Canfield1991,Bay2012b}, LaPtBi~\cite{Goll2008} and LuPtBi~\cite{Mun2010,Tafti2013}, are low carrier systems and the transport properties reveal semi-metallic behaviour. For LuPtBi, surface metallic states have been observed in ARPES experiments~\cite{Liu2011}, but solid evidence for a topological non-trivial state has not been provided to date. Interestingly, superconductivity has also been reported for LaPtBi ($T_c = 0.9$~K~\cite{Goll2008}) and LuPtBi ($T_c = 1.0$~K~\cite{Tafti2013}). The non-trivial topology of the electron bands makes these platinumbismuthides candidate for topological superconductivity, with mixed parity Cooper pair states in the bulk and protected Majorana surface states~\cite{Hasan&Kane2010,Qi&Zhang2010}. The field of topological superconductors attracts much attention, but unfortunately, hitherto, only a few candidate materials have been discovered. Other potential candidates are Cu$_x$Bi$_2$Se$_3$~\cite{Hor2010,Bay2012} and Sn$_{1-x}$In$_x$Te~\cite{Sasaki2012}. Finally, we mention that the related non-centrosymmetric Half Heusler compound ErPdBi was recently put forward as a new platform for the study of the interplay of  topological states, superconductivity ($T_c = 1.22$~K) and magnetic order (the N\'{e}el temperature $T_N = 1.06$~K)~\cite{Pan2013}.

\section{Previous work and motivation}

YPtBi crystallizes in a cubic structure with lattice constant $6.650$~{\AA} and space group $F\overline{4}3m$~\cite{Canfield1991}. Magnetotransport data taken on single crystals grown from Bi flux show semi-metallic~\cite{Butch2011,Canfield1991} or weak-metallic~\cite{Bay2012b} behaviour. This difference in transport behaviour is reflected in the hole carrier concentration, $n_h$, at liquid helium temperatures, which equals $2 \times 10^{18}$~cm$^{-3}$~\cite{Butch2011} and $2.2 \times 10^{19}$~cm$^{-3}$~\cite{Bay2012b}, respectively. The low carrier concentration results in a very small value of the Sommerfeld coefficient in the specific heat: $\gamma \leq 0.1$~mJ/molK$^2$~\cite{Pagliuso1999}. Magnetic susceptibility, $\chi$, measurements in an applied field of 0.5~T show YPtBi is diamagnetic with a temperature independent $\chi = -10^{-4}$ emu/mol~\cite{Pagliuso1999}.

Hitherto, the superconducting state of YPtBi has mainly been characterized by resistivity measurements (in field)~\cite{Butch2011,Bay2012b}. A sharp drop to zero resistance is observed at $T_c =0.77$~K. The upper critical field, $B_{c2}$ attains a value of 1.2 T in the limit $T \rightarrow 0$, which translates into a superconducting coherence length $\xi = 17$~nm. YPtBi is a clean-limit superconductor, with a mean free path $\ell \sim 100$~nm $> \xi$. The relatively clean nature of YPtBi is furthermore demonstrated by the observation of Shubnikov-de Haas oscillations~\cite{Butch2011} that yield an effective mass of $0.15 m_e$. Ac-susceptibility measurements, $\chi _{ac}$, show a diamagnetic signal below $T_c$ but the response is sluggish~\cite{Butch2011,Bay2012b}. Heat capacity, $C(T)$, measurements around the normal-to-superconducting transition~\cite{Pagliuso2013} do not show the universal step $\Delta C/ \gamma T_c \simeq 1.43$ expected for a weak coupling spin singlet superconductor, but rather a break in slope of $C/T$ at $T_c$. Thus the specific heat data fail to provide evidence for bulk superconductivity.

Electrical resistivity measurements under pressure show $T_c$ increases at a linear rate of $0.044$~K/GPa~\cite{Bay2012b}. The upper critical field $B_{c2}(T)$ curves taken at different pressures collapse in a reduced plot onto a single curve with an unusual linear variation and values that largely exceed the model values for a weak-coupling spin-singlet superconductor~\cite{Werthamer1966}. These $B_{c2}$ data point to the presence of an odd-parity Cooper pair component in the superconducting order parameter, in agreement with predictions for noncentrosymmetric and topological superconductors~\cite{Frigeri2004,Schnyder2012,Fu&Berg2010}.

In this paper we report on the low-field magnetic response of single crystals of YPtBi and determine $B_{c1}$ and the London penetration depth $\lambda$. In order to elucidate the sluggish response of the ac-susceptibility~\cite{Butch2011,Bay2012b} below $T_c$ we have conducted systematic $\chi _{ac} (T)$ measurements as a function of the ac-driving field, $B_{ac}$. For small values $B_{ac} \leq 0.001$~mT we observe a pronounced superconducting transition. This, together with dc-magnetization measurements, provides unambiguous evidence for bulk superconductivity. The lower critical field $B_{c1} (T \rightarrow 0)$ extracted from the $\chi _{ac}$-data is surprisingly small and amounts to 0.008~mT. The second sensitive probe we used to study the low field magnetic response is $\mu$SR. Muon spin rotation experiments in an applied transverse field enable us to determine the London penetration depth $\lambda = 1.6 \pm 0.2~ \mu$m. In order to investigate the presence of an odd-parity component in the superconducting order parameter we have conducted muon spin relaxation measurements in zero field. These put an upper bound of 0.04~mT on the spontaneous magnetic field associated with the mixed parity order parameter.

\section{Sample preparation and characterization}

A single-crystalline batch of YPtBi was prepared out of Bi flux. The tiny crystals had predominantly a pyramid-shape (edge size $\leq 1$~mm) with the basis aligned with the [111] direction. Powder x-ray diffraction confirmed the $F\overline{4}3m$ space group. The batch was further characterized by magnetotransport and ac-susceptibility measurements. The resistance, $R(T)$, measured on one of the tiny crystals reveals semi-metallic behaviour with a broad maximum around 80~K as shown in Fig.~1. Magnetoresistance traces taken at liquid helium temperatures display Shubnikov-de Haas (SdH) oscillations, which attests the high quality of the sample. The hole carrier concentration, $n_h$, deduced from the SdH signal equals $ 1.3 \times 10^{18}$~cm$^{-3}$. This number is in agreement with the value reported previously for a semimetallic sample~\cite{Butch2011}. The superconducting transition to zero resistance for this crystal is shown in the inset of Fig.~1. $T_c$ determined by the midpoint of the transition is $0.98$~K, which is higher than in the literature. The width of the superconducting transition is relatively large $\Delta T_c = 0.36$~K with a weak tail towards low temperatures. The diamagnetic $\chi_{ac}$ signal, measured at a frequency of 16~Hz and a driving field $B_{ac} = 0.026$~mT, sets in at $T=0.80$~K, \textit{i.e.} when the transition in the resistance is complete. Note the sluggish transition in $\chi_{ac}$ obtained in this way becomes much sharper when $B_{ac}$ is reduced to below $0.001$~mT (see the next section).

\begin{figure}
\includegraphics[width=7cm]{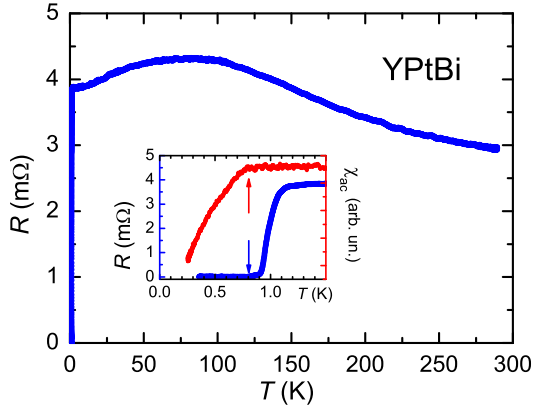}
%Fig_YPtBi_Resistance_and_chi_ac
\caption{(color online) Resistance of YPtBi as a function of temperature showing semi-metallic behaviour. Inset: Superconducting transition in resistance (blue symbols; left axis) and in ac-susceptibility in a driving field $B_{ac} = 0.026$~mT (red symbols; right axis); blue and red arrows indicate $R=0$ and the onset of the diamagnetic signal, respectively.}
\end{figure}

\begin{figure}
\includegraphics[width=8cm]{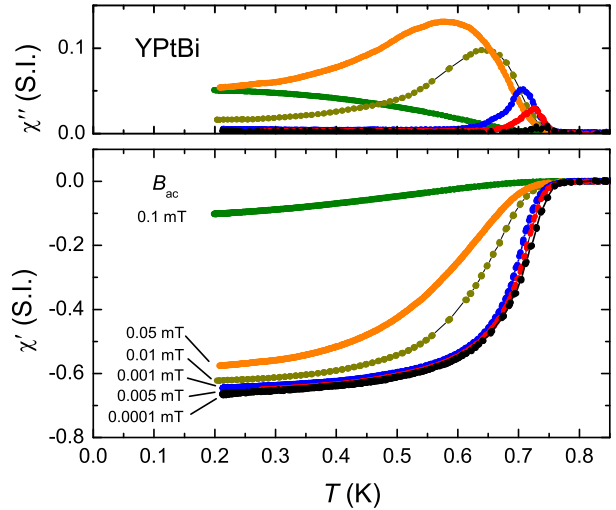}
%Fig_YPtBi_ac-susceptibility
\caption{(color online) Ac susceptibility as a function of temperature of YPtBi for different driving fields $B_{ac}$ as indicated. Lower frame: $\chi ^{\prime}$; upper frame: $\chi ^{\prime \prime}$. }
\end{figure}

\section{Low-field magnetization and ac susceptibility}

For the magnetic measurements 10 small single crystals were arranged in a circular cluster with a total mass of 42~mg. Dc-magnetization and ac-susceptibility measurements were made using a SQUID magnetometer, equipped with a miniature dilution refrigerator, developed at the N\'{e}el Institute. As concerns $\chi_{ac}$, the in-phase, $\chi ^{\prime}$, and out-of-phase, $\chi ^{\prime \prime}$, signals were measured in driving fields $B_{ac} \leq 0.1$~mT with a frequency of 2.1~Hz. The diamagnetic signal is corrected for demagnetization effects: $\chi_{diam} = -1/(1-N)$. Here we used $N=1/3$, since the sample is effectively a `powder'.

The temperature variation of the ac-susceptibility is reported in Fig.~2. For the collection of single crystals we find $T_c = 0.77$~K, as determined by the onset temperature of the diamagnetic signal. This value of $T_c$ is in good agreement with the results reported in Refs.~\cite{Butch2011,Bay2012b}. Note $\chi ^{\prime} (T)$ and $\chi ^{\prime \prime} (T) $ show a strong dependence on $B_{ac}$. For the smallest values of $B_{ac}$ ($\leq 0.001$~mT) the standard behaviour for a superconductor is observed: $\chi ^{\prime} (T)$ shows a relatively sharp drop below $T_c$, and $\chi ^{\prime \prime} (T)$ shows a peak due to dissipation. However, with increasing values of $B_{ac}$ the transition broadens rapidly. This explains the sluggish temperature variation of $\chi ^{\prime}$ measured with $B_{ac} = 0.026$~mT reported in Fig.~1. The strong variation as a function of $B_{ac}$ indicates a small value of the lower critical field $B_{c1}$. Another important result is the large value of the diamagnetic screening signal which is reached for $B_{ac} = 0.0001$~mT. This points to a superconducting volume fraction of 67 \%.

\begin{figure}
\includegraphics[width=8cm]{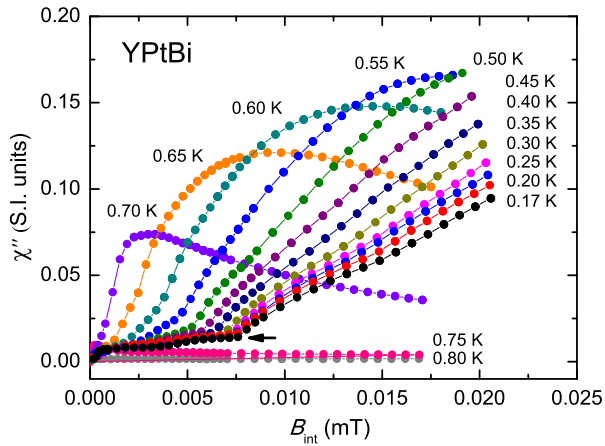}
%Fig_YPtBi_chi_doubleprime
\caption{(color online) $\chi ^{\prime \prime}$ as a function of the internal field $B_{int}$ at temperatures as indicated. $B_{int}$ is obtained by correcting for demagnetization effects: $B_{int} = B_{ac} (1-N \chi ^{\prime})$. The kink locates the lower critical field $B_{c1}$ as indicated by the arrow for $T = 0.17$~K.  }
\end{figure}

The ac-susceptibility signal measured as a function of $B_{ac}$ provides a very sensitive way to probe $B_{c1}$~\cite{Paulsen2012}. Notably, the imaginary part of the susceptibility, $\chi ^{\prime \prime}$, which is related to losses and hysteresis, is an excellent indicator of the first flux penetration in the sample. If there is a perfect Meissner state up to $B_{c1}$ then $\chi ^{\prime \prime} (T) = 0$ for $B_{ac} \leq B_{c1}$ and $\chi ^{\prime \prime} (T) = \beta (B_{ac}-B_{c1})/B_{ac}$ for $B_{ac} >  B_{c1}$. Here $\beta$ is a parameter that depends on the sample geometry and is related to screening currents according to the critical state model~\cite{Bean1964}. In Fig.~3 we report $\chi ^{\prime \prime}$ as a function of the internal field $B_{int}$.  At the lowest temperature, $T= 0.17$~K, the clear kink observed near 0.0076~mT locates $B_{c1}$. Upon increasing the temperature the kink becomes more and more rounded. $B_{c1} (T)$ determined in this way is traced in Fig.~4. In the normal state, \textit{e.g.} at $T= 0.80$~K, $\chi ^{\prime \prime} (T)$ is essentially flat. We remark that in the Meissner state $\chi ^{\prime \prime} (B_{int})$ is not equal to 0, but shows a weak quasi-linear increase. The origin of this behaviour is not clear. Possible explanations are sharp sample edges where flux could penetrate more easily, and the presence of an impurity phase with a very small critical field ($< 0.001$~mT). In the limit $T\rightarrow~0, B_{c1} = 0.0078$~mT. In Fig.~4 we also compare the $B_{c1}$-data with the standard BCS quadratic temperature variation (see caption Fig.~4). A clear departure is found at the lowest temperatures. Alternatively, $B_{c1}$ can be deduced from the dc-magnetization measured as a function of the applied field. $M (B_{appl})$-data taken at $T= 0.17$~K are shown in the inset of Fig.~4. $B_{c1}$ determined in this way amounts to 0.0083 mT, in good agreement with the method described above.

\begin{figure}
\includegraphics[width=8cm]{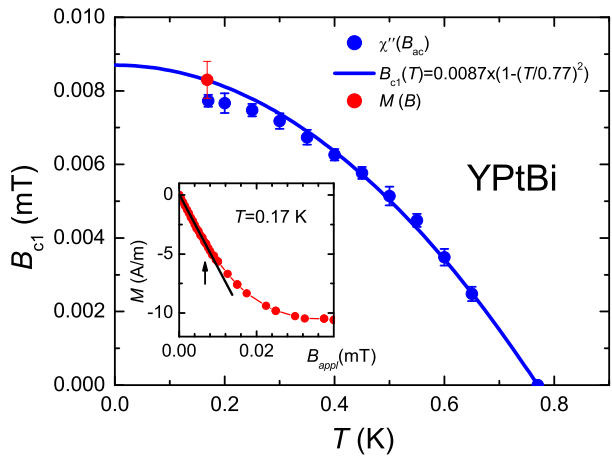}
%Fig_YPtBi_Lower_critical_field
\caption{(color online) The lower critical field $B_{c1}$ as a function of temperature. The solid line represents a quadratic dependence $B_{c1}(T) = B_{c1}(0)(1-(T/T_c)^2)$ with $B_{c1}(0)=0.0087$~mT and $T_c = 0.77$~K. Inset: Dc-magnetization versus applied field at $T =0.17$ K. The black arrow indicates where  $M(B_{appl})$ deviates from linear behaviour (black straight line) and flux penetrates the sample.}
\end{figure}

\begin{figure}
\includegraphics[width=8cm]{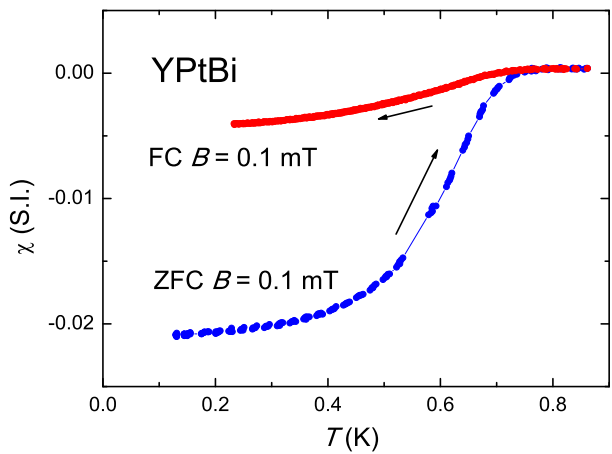}
%Fig_YPtBi_Flux_expulison
\caption{(color online) DC-susceptibility versus temperature in an applied field of 0.1~mT. After cooling in $B=0$ (ZFC), a magnetic field of 0.1 mT is applied. Next the sample is heated to above $T_c$ and subsequently cooled in 0.1~mT (FC) to demonstrate flux expulsion. }
\end{figure}

Finally, we present in Fig.~5 dc-magnetization measurements that provide solid evidence for bulk superconductivity. After cooling in zero field, a field of $0.1$~mT is applied in the superconducting state. This gives rise to the diamagnetic screening signal. Upon heating the sample to above $T_c$, the diamagnetic signal vanishes. On subsequent cooling, flux expulsion is clearly observed, which corresponds to a Meissner fraction of 0.4~volume~\%. Note this fraction is very small, because the applied field is much larger than $B_{c1}$ and flux pinning is strong (see also Fig.~2).

\section{Muon spin rotation and relaxation}

Muon spin rotation and relaxation experiments ($\mu$SR) were carried out at the $\pi$M3 beam line at the Paul Scherrer Institute. The motivation for the experiments was two-fold: (\textit{i}) to investigate the appearance of a spontaneous magnetic signal due to the breaking of time reversal symmetry associated with an odd parity component of the superconducting order parameter, and (\textit{ii}) to determine the London penetration depth, $\lambda$, in the superconducting state. Measurements were made in the Low Temperature Facility (LTF) in the temperature range $T=0.02-1.8$~K in zero field (ZF) and weak transverse fields (TF). The `polycrystalline' sample consisted of a large ensemble of tiny crystals glued in a random crystal orientation on a silver backing plate with General Electric (GE) varnish to ensure good thermal contact. The sample area amounted to $10 \times 14$ mm$^2$. Ac-susceptibility measurements confirmed $T_c$ = 0.77 K.

\begin{figure}
\includegraphics[width=8cm]{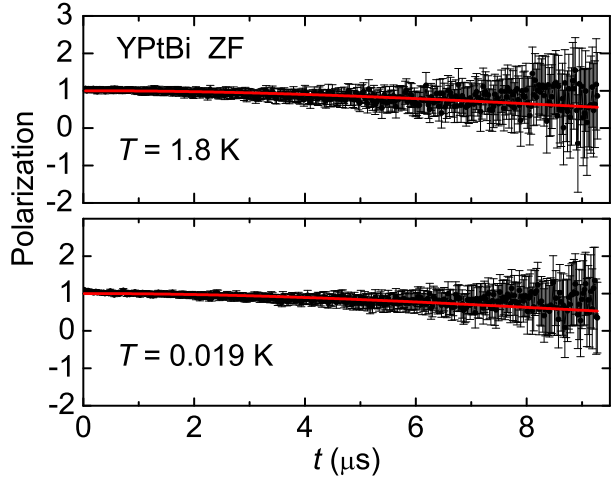}
%Fig_YPtBi_Lower_critical_field
\caption{(color online) Time dependence of the normalized muon spin depolarization of YPtBi in zero field at 1.8 K (upper frame) and 0.019 K (lower frame). The solid red lines are fits to the Kubo-Toyabe depolarization function eq.1. }
\end{figure}

\begin{figure}
\includegraphics[width=8cm]{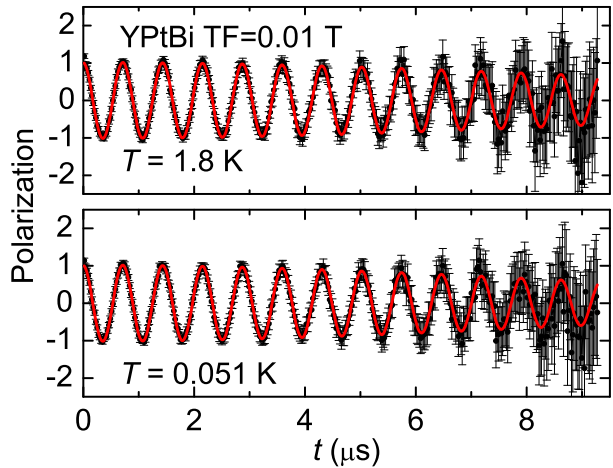}
%Fig_YPtBi_Lower_critical_field
\caption{(color online) Time dependence of the normalized muon spin depolarization of YPtBi in a transverse field of 0.01~T at 1.8 K (upper frame) and 0.051 K (lower frame). The solid red lines are fits to a depolarization function with a precession frequency and Gaussian damping, eq.2.  }
\end{figure}

In Fig.~6 we show ZF $\mu$SR spectra taken at 1.8~K and 0.019~K. The depolarization of the muon ensemble is weak and does not change significantly with temperature. The spectra are best fitted with the standard Kubo-Toyabe function:

\begin{equation}
G_{KT} (t)= \frac{1}{3} + \frac{2}{3} (1- \Delta_{KT} ^{2} t^{2})
\exp (- \frac{1}{2} \Delta_{KT} ^{2} t^{2} )
\end{equation}

The Kubo-Toyabe function describes the muon depolarization due to an isotropic Gaussian distribution of static internal fields centered at zero field. $\Delta_{KT} = \gamma_{\mu} \sqrt{\langle
B^2 \rangle}$ is the Kubo-Toyabe relaxation rate, with $\gamma_{\mu}$ the muon gyromagnetic ratio ($\gamma_{\mu} /2 \pi = 135.5~$MHz/T) and $\langle B^2 \rangle$ the second moment of the
field distribution. Note the characteristic minimum at $t/\Delta_{KT} \simeq 1.74$ and the recovery of the 1/3 term is not observed in this time window because of the small relaxation rate. In this temperature range the extracted values of $\Delta_{KT}$ are the same within the error bars. The average value is $0.129 \pm 0.004~\mu$s$^{-1}$ (see Fig.~8). The field distribution is most likely arising from the nuclear moments of the $^{89}$Y, $^{195}$Pt and  $^{209}$Bi isotopes which can be considered as static within the $\mu$SR time-window. The  sizeable value of $\Delta_{KT}$ reflects a broad distribution of internal fields, which can be attributed to the polycrystalline nature of the sample. The uncertainty in $\sigma _{KT}$ allows to determine an upper bound for a possible additional spontaneous magnetic field below $T_c$ of 0.04~mT.

$\mu$SR spectra in a TF $B_{TF} =0.01$~T taken at 1.8 K and 0.051 K are shown in Fig.~7. Note the TF was applied after cooling in ZF. The spectra were fitted to the depolarization function:

\begin{equation}
G (t)= \exp (-\frac{1}{2} \sigma_{TF}^2 t^2) \cos (2\pi\nu t+\phi)
\end{equation}

Here $\sigma _{TF}$ is the Gaussian damping factor, $\nu = \gamma_{\mu} B_{TF} /2 \pi$ is the frequency where $B_{\mu}$ is the average field seen by the muon ensemble and $\phi$ is a phase factor. The temperature variation of $\sigma _{TF}$ is shown in Fig.~8. In the normal phase $\sigma _{TF}=0.105 \pm 0.005 ~ \mu$s$^{-1}$ and represents here again a field distribution due to the nuclear moments. Upon lowering the temperature a weak increase is found below $T_c$ that is attributed to the $\mu ^+$ depolarization  $\sigma_{FLL}$ due to the flux line lattice. The corresponding relaxation rate can be calculated from the relation $\sigma_{FLL}=(\sigma^2_{TF, T<T_c} - \sigma^2_{TF, T>T_c})^{1/2}$ and is estimated at $0.04\pm 0.01 ~\mu$s$^{-1}$. For a type II superconductor and $B \gg B_{c1}$ the London penetration depth can be estimated from the relation ${\langle
B^2 \rangle}=0.003706 \times \Phi_0 ^2 / \lambda^4$, where $\Phi_0$ is the flux quantum~\cite{Brandt1988}. With $\sigma_{FLL}=0.04 \pm 0.01 ~\mu$s$^{-1}$ we calculate $\lambda= 1.6 \pm 0.2 ~\mu$m for $T \rightarrow 0$. This value is about three times larger larger than the lattice parameter of the trigonal flux line lattice induced by the 0.01~T field: $a_{\triangle}=(4/3)^{1/4}(\Phi_0 /B)^{1/2} = 0.49~\mu$m. Note the relatively large error bar on $\sigma_{FLL}$ does not allow for an accurate determination of the temperature variation of $\lambda$, which impedes the detection of possible power laws in the excitation spectrum of the superconducting state.

\begin{figure}
\includegraphics[width=8cm]{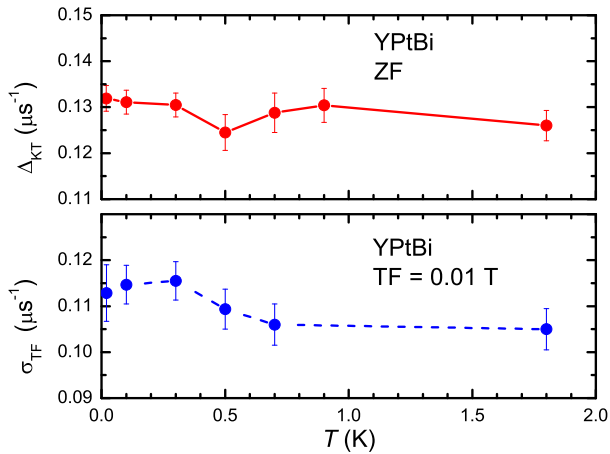}
%Fig_YPtBi_Lower_critical_field
\caption{(color online) Temperature variation of the Kubo-Toyabe relaxation rate in zero field (upper frame) and of the Gaussian damping rate in a transverse field of 0.01 T (lower frame). Solid lines connect the data points. }
\end{figure}

\section{Discussion}

Since YPtBi is a low-carrier system the London penetration depth is expected to be large. The London penetration depth is related to the superfluid density $n_s$ via the Ginzburg-Landau relation $n_s = m_{h}^{*}/\mu _0 e^2 \lambda^2$, where $m_{h}^*$ is the effective mass of the charge (hole) carriers, $\mu_0$ is the permeability of the vacuum and $e$ the elementary charge. With the experimental values $\lambda= 1.6 \pm 0.2 \mu$m and $m_h ^* = 0.15~ m_e$~\cite{Butch2011} we calculate $n_s = (1.7 \pm 0.4) \times 10^{18}$~cm$^{-3}$, which is in agreement with the carrier concentration determined from the SdH effect, $n_s = 1.3 \times 10^{18}$~cm$^{-3}$ (see section 3). Thus the transport and the TF $\mu$SR data give a consistent picture.

With help of the characteristic lengths of the superconducting state, $\lambda$ and $\xi$, the lower critical field can be deduced from the Ginzburg-Landau relation $B_{c1}=\Phi_0 \ln (\lambda / \xi)/( 4 \pi \lambda ^2 )$. Here $\xi$ = 17~nm~\cite{Bay2012b} is calculated from the upper critical field $B_{c2} = \Phi _0 / 2 \pi \xi ^2$ and the Ginzburg-Landau parameter $\kappa = \lambda / \xi \simeq 94$. Using the experimental value $\lambda= 1.6 \pm 0.2 ~\mu$m we obtain $B_{c1} = 0.29 \pm 0.05$~mT ($T \rightarrow 0$). Surprisingly this value is a factor 36 larger than the measured value of 0.008~mT (section 4). Or, the other way around, to match the measured $B_{c1}$-value $\lambda$ should be equal to $\sim 11 ~\mu$m. This in turn would entail $ \sigma_{FLL} \sim 0.001~\mu$s$^{-1}$, a value not compatible with the analysis of the $\mu$SR and transport data. It is tempting to attribute this discrepancy to an intricate relation, beyond the simple Ginzburg-Landau approach, between $B_{c1}$ and $\lambda$. In particular a non-unitary Cooper pair state will have an intrinsic magnetic moment, which could result in a very small $B_{c1}-$value~\cite{Paulsen2012}. The spontaneous internal field at the muon localization site associated with the unitary state should however be smaller than 0.04~mT in the limit $T \rightarrow 0$. Further evidence for an odd-parity component in the superconducting order parameter is provided by the reduced $B_{c2}-$values~\cite{Bay2012b}. We note that $B_{c1} (T)$ deviates from the standard BCS behaviour (Fig.~4) in the same temperature range as $B_{c2} (T)$. Clearly, this calls for theoretical studies as regards flux penetration in mixed order parameter component superconductors. From the experimental side, notably $\mu$SR, it will be highly desirable to work with a large homogeneous single crystal, which is expected to significantly reduce the background relaxation times. Together with improved statistics, this will enable one to resolve the temperature variation of $\lambda$ and to shed further light on the magnitude of the spontaneous internal magnetic moment.

\section{Summary}

The superconducting properties of YPtBi deserve ample attention because the crystal structure lacks inversion symmetry, which may give rise to unconventional superconductivity. Moreover, YPtBi has an electronic band inversion and is predicted to harbor topological surface states. We have investigated the low-field magnetic response by means of magnetization and $\mu$SR experiments.  Ac-susceptibility and dc-magnetization measurements provide unambiguous proof for bulk superconductivity. The lower critical field $B_{c1} = 0.008$~mT $(T \rightarrow 0)$ is surprisingly small. This is a robust property, which possible finds an explanation in a non-unitary superconducting order parameter. Muon spin rotation experiments in a transverse magnetic field of 0.01~T show a weak increase of the Gaussian damping rate $\sigma_{TF}$ below $T_c$, which yields a London penetration depth $\lambda = 1.6 \pm 0.2~ \mu$m. The zero-field Kubo-Toyabe relaxation rate $\sigma_{KT}$ equals $0.129 \pm 0.004$~s$^{-1}$ and does not show a significant change below $T_c$. This puts an upper bound of 0.04~mT on the spontaneous magnetic field associated with a possible odd-parity superconducting order parameter component.

\vspace {10 mm}
\noindent
Acknowlegdments: This work is carried out in the research programme on Topological Insulators of the Foundation for Fundamental Research on Matter (FOM), which is part of the Netherlands Organisation for Scientific Research (NWO). Part of the experiments were carried out at the Swiss Muon Source S$\mu$S (PSI, Villigen, Switzerland) with support by the European Commission under the 7th Framework Programme through the Research Infrastructures action of the Capacities Programme, NMI3-II Grant number 283883. Work at Brookhaven National Laboratory was carried out under the auspices of the US Department of Energy, Office of Basic Energy Sciences under Contract No. DE-AC02-98CH1886.

\clearpage

\bibliographystyle{elsarticle-num}
\bibliography{RefsTI}

\end{document}